\documentclass{article}
\usepackage{breckenr2005}
\usepackage{graphicx}
\frompage{000} \topage{000}                                              
\def\rightmark{The future RHIC-II program }
\def\leftmark{R. Bellwied}

\title{Hadronization and chirality in strongly interacting partonic matter - the future of the RHIC program}
\authors{
{Rene Bellwied}\\[2.812mm]
{\normalsize
Department of Physics and Astronomy, Wayne State University \\
Detroit, MI 48201, USA
}}

\abstract{New physics and detector concepts for a future pp and
heavy ion program at the RHIC-II accelerator facility will be
discussed. I will focus on hadronic observables which enable us to
gain a better understanding on the hadronization from a sQGP and the
chiral symmetry restoration in a sQGP. The ultimate question of how
matter acquires mass can be addressed by this program in a
complementary way to the Higgs search in high energy physics. The
contributions of the RHIC program to the study of QCD will be
discussed in detail.}


\begin{document}

\maketitle
\setcounter{page}{1}

\section{Introduction}


Over the past four years the ongoing RHIC-I program has attained an
impressive set of data. The experimental findings and conclusions
were summarized in four independent white paper documents by the
four RHIC experiments \cite{whitepapers}. All whitepapers concluded
that a high density partonic matter state was formed, which exhibits
very strong coupling. The collectivity of the state seems well
described by hydrodynamical models assuming a very low viscosity
\cite{teaney}. In other words the system behaves more like an ideal
fluid than a weakly interacting electromagnetic plasma, best shown
by the good hydrodynamical description of the kinematic properties
of the low momentum bulk matter formed in the interaction, in
particular the elliptic and radial flow as a function of centrality
\cite{v2},\cite{radial}. The strong mass dependencies of both
measurements demonstrate the collectivity of most of the produced
matter, i.e. the 99.5\% of the particles with a transverse momentum
of less than 2 GeV/c. The very small mean free path assumed in the
hydrodynamical models hints at a rather strong coupling and large
interaction rate among the relevant degrees of freedom, at least
near the critical temperature. Furthermore the strength of the
elliptic and radial flow can only be reproduced when one assumes a
partonic equation of state for the early phase of the system
evolution \cite{huovinen}. In these detailed calculations it was
shown that about 80\% of the collective flow is due to early
partonic interactions, thus a purely hadronic phase can not account
for most of the collective behavior. This is only an indirect and
model dependent proof of a partonic state, though.

A more direct measure of the partonic nature of the new state is
given by the interaction of hard probes, produced in the same
collision, with the medium, in particular through measurements of
nuclear suppression factors R$_{AA}$, elliptic flow v2, and
di-hadron correlations $\Delta\phi$ at high pt
\cite{v2},\cite{star-raa},\cite{star-corr}. The nuclear suppression
factors, i.e. the ratio of the particle momentum spectra in AA
collisions compared to properly scaled pp collision spectra, shows
that in AA collisions the spectrum is quenched at high pt. A
possible explanation of this effect is that in the dense partonic
medium the fragmenting partons lose large amounts of energy due to
radiative energy loss, i.e. gluon bremsstrahlung \cite{xn-wang}. By
comparing measurements of the suppression factor in central AuAu
collisions at RHIC to central eA collisions at DESY, one can
conclude that the energy loss in matter produced at RHIC is about 15
times larger than in cold nuclear matter \cite{wang-wang}. The only
possible explanation is the formation of matter with partonic
degrees of freedom. This assumption is corroborated by the
measurement of jet quenching through di-hadron correlations
\cite{star-corr}. The away-side jet in a back-to-back di-hadron jet
needs to traverse the medium and is apparently quenched in central
AA collisions compared to pp collisions or peripheral AA collisions.

Finally the detailed particle identified measurement of elliptic
flow and nuclear suppression factors at high pt seems to shed some
light on the question of the nature of the partonic degrees of
freedom above the critical temperature. In both measurements an
interesting scaling has been found, which can be explained by
assuming constituent quarks as the relevant degrees of freedom
\cite{star-ident}, \cite{fries}. There is a distinct baryon/meson
difference in the kinematics which are dominated by patron
interactions, i.e. the v2 and the particle production above two
GeV/c. Furthermore this baryon/meson difference seems to disappear
at around six GeV/c which might signal the dominance of
fragmentation as the main production mechanism at higher transverse
momentum. Between two and six GeV/c the mechanism though seems to be
recombination of constituent quarks, i.e. colored objects with a
finite mass. Recombining quarks will not fragment. They need to be
generated through strings, though, which means in order to have a
pool of thermal partons to recombine, most strings have to dissolve
during early times of the system evolution. This is most likely a
pt-dependent effect, i.e. lower momentum strings will dissolve
whereas at higher momentum string fragmentation will still dominate.

In summary the RHIC-I measurements establish a new phase of matter,
but the properties of this phase are unexpected and require further
investigation, in particular because they seem to be sensitive to
the hadronization mechanism from the plasma to the hadronic phase.
This might enable us to answer a central question of QCD, namely how
do hadrons attain their mass and how does the strong coupling evolve
from the initial state down to the critical temperature. In the
following I will discuss an experimental program which addresses the
question of hadronization and chiral symmetry by comparing detailed
particle identified measurements in pp and AA collisions at RHIC-II.

\section{Why is pp so important ?}

In order to understand the modification of the basic hadronization
process in AA collisions we need to understand fragmentation in
elementary collisions. This was a topic in high energy physics for
many decades which culminated, from a theory point of view, in the
so-called factorization theorem \cite{fact1}, \cite{fact2}.
Factorization means that in order to calculate the actual hadron
production cross section in elementary collisions one needs to
parametrize additional terms besides the hard parton cross section
which can be calculated from first principles. The parametrization
of a.) the initial parton distribution function PDF and b.) the
fragmentation function FF is based on experimental results. These
contributions to a next-to-leading order (NLO) perturbative QCD
calculation have been studied in detail in the past \cite{hera-pdf},
\cite{kkp}. One conclusion was that fragmentation is universal, i.e.
a FF parametrized for e$^{+}$e$^{-}$ collisions can be applied to pp
collisions \cite{kkp}. This rule proved to be correct also at RHIC
energies, where the early neutral pion measurements by PHENIX in pp
collisions were very well described by FF deduced from e+e-
experiments according to Kraemer, Kniehl and Poetter (KKP)
\cite{phenix-pp}. When applied to higher mass mesons and baryons,
though, the original parametrization failed, as shown in the
comparison to preliminary STAR strange particle data \cite{heinz-pp}
on Fig.1.

\begin{figure}
\begin{center}
\includegraphics[width=3.5in]{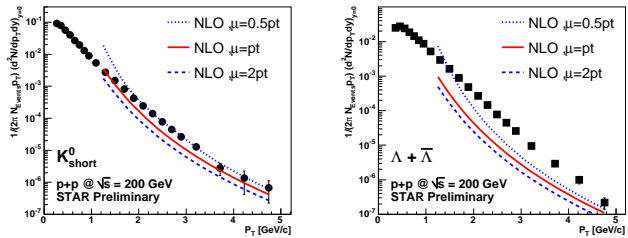} \caption{K0 and
$\Lambda$ particle spectra measured by the STAR collaboration in pp
collisions. The spectra are compared to NLO calculations by W.
Vogelsang using fragmentation functions by Kniehl et al. \cite{kkp}
and deFlorian et al. \cite{deflorian}.} \label{fig:heinz-pp}
\end{center}
\end{figure}

In a recent paper by Albino, Kraemer and Kniehl (AKK), it was shown
that the RHIC data can be better described when quark separation in
the FF is used, i.e. the contribution of each quark flavor to the
final hadron is computed separately and then integrated to the final
FF \cite{akk}. Interestingly the study by AKK on the strange mesons,
and independently by Bourrely and Soffer on the strange baryons
\cite{bourelly}, shows that there are considerable contributions
from fragmentation of non-valence quarks in the production cross
section, in particular at low fractional momentum z. Fig.2 shows the
quark separated fragmentation function for octet baryons in
elementary e$^{+}$e$^{-}$ collisions at $\sqrt{s}$ = 91.2 GeV. Fig.3
shows a comparison of the AKK calculation using quark separated
fragmentation functions to K$^{0}_{s}$ spectra from STAR and UA1.

\begin{figure}
\begin{center}
\includegraphics[width=2.5in]{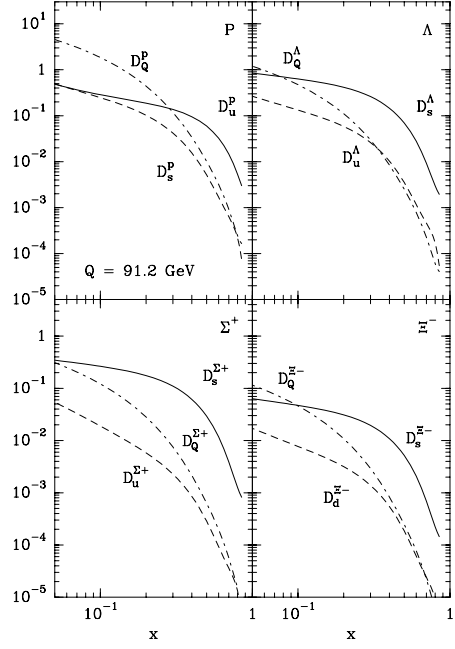} \caption{Contributions of individual quark
fragmentation functions to octet baryon production at
$\sqrt{s}$=91.2 GeV in e$^{+}$e$^{-}$ collisions according to the
statistical approach by Bourrely and Soffer \cite{bourelly}.}
\label{fig:bourrely}
\end{center}
\end{figure}

\begin{figure}
\begin{center}
\includegraphics[width=2.5in]{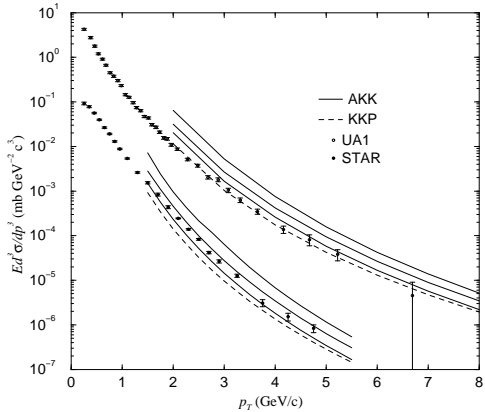} \caption{STAR and UA1 K0 spectra compared
to NLO calculations using quark separated fragmentation functions.
Details of the calculation are explained in \cite{akk}.}
\label{fig:akk}
\end{center}
\end{figure}

One can conclude that the fragmentation process even in pp is more
complex than originally thought, and measuring the particle
identified yields of heavy particles in pp collisions at RHIC-II out
to very high momentum is of utmost importance. In order to proof a
particular quark contribution to a final hadron yield one might then
be able use the medium modification of this traversing parton in the
sQGP formed in AA collisions, as will be explained in the following
section. One important point to keep in mind in the discussion of pp
collisions is that a recombination model, which does not assume
thermalized partons, but rather coalesces string partons with low
momentum partons, can also describe the pp data \cite{hwa-old}, and
in particular can describe certain measurements in elementary
collisions such as leading particle asymmetries, which can not be
explained by simple string fragmentation \cite{e791},
\cite{asymm-reco}.

\section{Measuring medium modification in AA.}

In order to determine a specific parton contribution to the hadronic
cross section one might be able to use the medium modification of
the fragmentation in AA collision, if one can show that different
partons exhibit differing energy loss in the quenching medium. Most
of the presently applied models assume that the energy is lost
through gluon radiation, and they treat the partonic energy loss
quite universal, i.e. no difference between parton flavors is
assumed \cite{gyulassy}. Recently though it was shown that in order
to describe the large v2 at high pt it might be necessary to assume
that there is also an enhanced elastic partonic cross section which
leads to a stronger contribution from collisional energy loss
\cite{molnar}. This cross section will vary as a function of the
parton momentum and the parton mass. In order to describe the high
pt v2 this model overestimates the nuclear suppression factor,
though. So a common description of R$_{AA}$ and v2 at high pt is
still an unresolved problem. Studying the nature of the energy loss
in the medium in detail (collisional vs. radiative, quark vs. gluon,
light vs. heavy quark, energy and density dependence, RHIC vs LHC)
will require to measure the R$_{AA}$ out to the highest pt, i.e. as
close to the kinematic limit as possible using identified particles
\cite{vitev}. But even in the theory of purely radiative energy
loss, differences between heavy quarks and light quarks (dead cone
effect \cite{kharzeev-dead}) and quarks and gluons (non-abelian
energy loss \cite{non-abelian}) have been postulated in theory and
are presently being investigated by the RHIC-I experiments, thus it
is probably save to assume that it will be very unlikely that the
medium modification of fragmentation will be universal among all
partons. Therefore, by comparing the energy loss for different
fractional momenta z, we might be able to determine the
probabilities of each parton contributing to each basic hadronic
fragmentation function. In order to do this measurement at the
required level of detail we need to be able to determine the
fractional momentum of the produced hadron directly. An unambiguous
z measurement can be accomplished in $\gamma$-jets, where the photon
forms one jet in a di-jet event. The direct photon will not fragment
and thus carries the full jet energy, which is then equivalent to
the full jet energy of the hadron jet on the away-side. Thus the
fractional momentum for any particle in the hadron jet can be
determined by measuring the energy of the direct photon. Clearly in
order to draw any conclusions from the $\gamma$-jet measurement it
is important to measure the away-side hadrons in coincidence with
the direct photon, and it is the acceptance window of the experiment
for the away-side hadron jet which limits the $\gamma$-jet
measurements at RHIC-II. A detailed simulation of the relative
opening angles in gamma-jet events is shown in \cite{harris-here}.
Table 1 shows the relevant yields in our proposed R2D detector per
RHIC-II year.

\begin{table}[hb]
\vspace*{-12pt} \caption[]{Di-jet and $\gamma$-jet yields in a
RHIC-II year, assuming 30 nb$^{-1}$ and 50\% duty
cycle.}\label{tab1} \vspace*{-14pt}
\begin{center}
\begin{tabular}{lll}
\hline\\[-10pt]
Condition & Number of events \\
\hline\\[-10pt]
40 GeV di-jets & 120,000 \\
$\gamma$-jets with E$_{\gamma}$ = 10 GeV and proton above 5 GeV/c &
2,000,000 \\
$\gamma$-jets with E$_{\gamma}$ = 15 GeV and proton above 5 GeV/c &
200,000 \\
$\gamma$-jets with E$_{\gamma}$ = 20 GeV and proton above 5 GeV/c &
19,000 \\
\hline
\end{tabular}
\end{center}
\end{table}

These measurements are statistics limited because one requires a
coincidence between an identified high pt hadron on the hadron jet
side and the direct photon on the $\gamma$-jet side. In order to
make an unambiguous fragmentation measurement we also need to
consider that the intermediate pt range might be 'contaminated' by
recombination processes. Therefore the range above 6 GeV/c for the
identified hadron would be ideal. In order to push the fractional
momentum coverage to low z, this requires to measure $>$15 GeV/c
direct photons.

\section{What can AA measurements tell us about degrees of freedom
above T$_{c}$ ?}

One key question that arises from the RHIC-I experimental results is
the origin of the rather strong coupling that allows hydrodynamics
to describe the collective properties of the phase. Indirect
measurements of the initial energy density through transverse energy
measurements under the assumption of early thermalization show that
the initial temperature achievable at RHIC is about 2 T$_{c}$
\cite{phenix-energy}. Is the sQGP between 1-2 T$_{c}$ really a
surprise ? Lattice QCD calculations show us that the Stefan
Boltzmann limit of an ideal gas is not reached at these temperatures
\cite{lattice}. The lattice properties at 2 T$_{c}$ undershoot the
ideal gas limit by about 15\%. If one believes that these 15\% can
cause the strong coupling we see at RHIC, then the question is
justified whether the ideal gas limit can be reached at higher
initial temperature, which could cause a difference in the
properties of the QGP measured at RHIC-II and LHC. Theoretical
guidance on this topic is mixed with certain theories showing a
large difference between the coupling at 1.5 Tc (RHIC) and 4 T$_{c}$
(LHC) \cite{peshier}, \cite{pisarski}, whereas most lattice
calculation show almost no difference \cite{lattice}. The same
lattice calculations, though, show that the strong coupling constant
is running as a function of temperature and distance, and it seems
even in these calculations that the coupling strength picks up
considerably below 2 T$_{c}$ \cite{kaczmarek}. We therefore might
have a detailed difference in plasma behavior at the LHC and RHIC.
Any measurement that could proof such a difference will require
sensitivity to the very early time of the system evolution. One can
speculate that the elliptic flow which is due to the original
eccentricity of the system at time zero might be reduced if the
system has time to get back to a more spherical shape in a weakly
interacting system before the thermalized sQGP phase is reached.
Therefore the v2 would decrease from RHIC to LHC energies.

In a lattice QCD calculation \cite{kaczmarek}, as well as in other
calculations such as a quasi-particle picture above T$_{c}$
\cite{peshier}, the interaction strength picks up when one
approaches T$_{c}$ from above, simply by having an interaction mass
or thermal mass produced through gluonic fields. There is no
explicit mass term attached to the partonic degree of freedom, which
is different from the constituent quark picture \cite{greco}. In
other words, simple lattice or quasi-particle pictures do not yield
constituent quarks above T$_{c}$, but simply lead to a one to one
duality between initial parton and final hadron. Therefore they
should not support a constituent quark scaling law for properties
developed above T$_{c}$. Another picture of degrees of freedom,
causing strong coupling above T$_{c}$, is the model of gluonic bound
states, according to Shuryak et al. \cite{shuryak}. In the model
these states will be melted at the initial T achievable at LHC, but
they could exist in the RHIC environment. They are mostly colored
but the few color neutral combinations could potentially be measured
as high mass resonance states. Again it is not clear how these
degrees of freedom can be reconciled with constituent quarks. In
summary there is no consistent dynamic picture of mass generation,
which evolves from an initial state (potentially the Color Glass
Condensate (CGC) \cite{mclerran}) through a thermalized massless
parton state through a constituent quark state to the final hadronic
state. Parts of this evolution could be tested at RHIC-II by
measuring probes such as energy loss and v2 which are sensitive to
the degrees of freedom above T$_{c}$. In particular, particle
identification will allow us to determine flavor and coupling
strength dependencies above T$_{c}$.

Another interesting question is whether constituent quarks
explicitly break chiral symmetry, because in recombination models
the valence quarks simply are assigned their actual constituent
mass, e.g. 300 MeV/c$^{2}$ for the light quarks, in order to
calculate the baryon/meson differences. The question of chiral
symmetry restoration in general has not yet been successfully
addressed in the RHIC-I program. The main measurement tools are
based on results from lower energies at SIS and the SPS
\cite{sis},\cite{ceres}. Here it is shown that masses at or below
production threshold are medium modified at relativistic energies,
but this seems an effect purely caused by a large hadronic medium
with finite baryochemical potential. Measurements in the Kaon system
at SIS and the vector meson spectrum at lower SPS energies have been
repeated at RHIC, but there are as of yet no results that hint at
chiral symmetry restoration in the partonic medium. The PHENIX
$\phi$ to KK measurement is probably the best resolution measurement
of this channel and it shows neither width nor mass shifts of the
vector meson \cite{phenix-phi}. The complementary $\phi$ to
e$^{+}$e$^{-}$ is still under investigation by PHENIX in order to
determine any modification to the branching ratio. In order to
better address the issue of chirality in the parton phase,
measurements of chiral partners might have to be obtained. The
shifting of chiral partners to the same mass or the melting of
chiral partners at the same T$_{c}$ would be indications of chiral
symmetry restoration $\cite{rapp}$. Unfortunately most chiral
partner systems are difficult to measure. The main candidate systems
are the $\sigma$ and the $\pi$ or the $\rho$ and the a$_{1}$. In
particular the latter is experimentally interesting \cite{delphi}.
Early measurements by STAR on the $\rho$ have proven to be
successful in establishing the signal above the hadronic cocktail
background \cite{fachini}. The question for RHIC-II will be whether
the a1 can be measured. The main problem is the large width of this
resonances, but its $\pi$+$\gamma$ decay channel is quite unique
assuming a good photon measurement in the intermediate pt range can
be performed. A new detector at RHIC-II might enable us to do so.

\section{Required detector capabilities of a new RHIC-II detector.}

The discussion of the necessary physics measurements in the previous
chapters leads to the following conclusions regarding a novel
experimental setup for RHIC-II:

a.) the detector needs to be as hermetic as possible, i.e. the
pseudo-rapidity needs to be 2$\pi$ radially and extended to the
largest pseudo-rapidity azimuthally. Existing high energy barrel
detectors reach out to $\eta$ = $\pm$3.5 and are used as a basis for
the proposed detector concepts. A specific forward program which
requires to extend the coverage out to about $\eta$ = $\pm$5 is not
discussed here but could be accommodated via a dedicated forward
spectrometer with its own magnetic field as described in the R2D
Expression of Interest \cite{r2d-eoi}.

b.) particle identification over the same hermetic coverage needs to
extend out to 20-30 GeV/c in order to enable our measurements of
fragmentation properties in pp and AA.

c.) the direct photon measurements in AA require a very high
resolution electromagnetic calorimeter (ideally a crystal
calorimeter), plus potentially hadronic calorimetry for the neutral
hadron energy component in the jet and in order to suppress
fragmentation photons via isolation cuts. A crystal calorimeter will
also potentially allow the measurement of decay photons from the a1
resonance.

d.) all detector components including the tracking, vertexing and
muon detectors need to be fast in order to serve as triggering
devices for rare processes and in order to obtain large statistics
minimum bias samples for the bulk processes.

e.) finally the detector requires a high magnetic field over the
central seven units of pseudo-rapidity in order to acquire the
necessary resolution of high pt tracks and in order to sweep a
considerable part of the bulk matter out of the mid-rapidity range
where most of the jet reconstruction analysis will be performed.

These five major points led to the design of R2D. Presently two
options for such a detector are being discussed, a large L-R2D based
on the SLD magnet dimensions, and a compact S-R2D based on the CDF
magnet dimensions. The CDF magnet is near identical in inner radius
and field strength to the CLEO and BABAR magnets and thus components
of all three of these detectors might be re-used upon their
decommissioning. L-R2D is shown in John Harris' contribution to this
workshop \cite{harris-here}. One possible variant of S-R2D is shown
in Fig.4.

\begin{figure}[hbtl]
\begin{center}
\includegraphics[width=4.in]{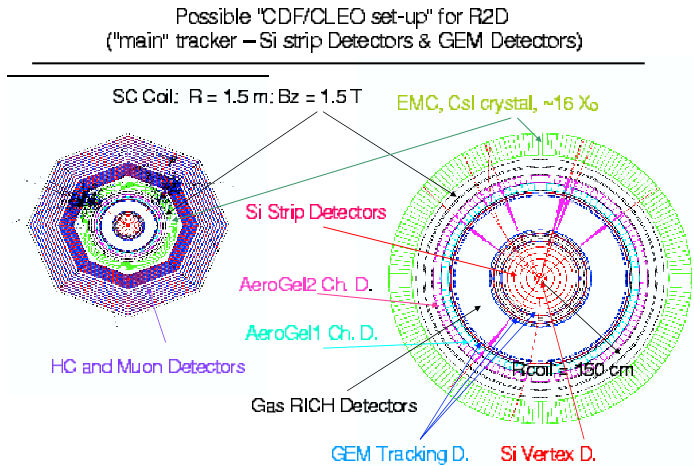}
\vspace{-4.5cm} \caption{A variant of the compact layout for a new
RHIC-II detector (S-R2D) based on a solid state tracking device and
RICH-type particle identification detectors.} \label{fig:nikolai}
\end{center}
\end{figure}

The L-R2D option is preferred because it features a sufficiently
large volume inside the magnet to allow for radially deep particle
identification detectors based on RICH technology. Unfortunately
L-R2D might require the construction of an additional experimental
area at RHIC, which will increase the total cost. The S-R2D is less
expensive (by about 40\%), but it is not yet determined how much the
restrictions in the particle identification capabilities compromise
the physics program. The variant of S-R2D shown in Fig.5 features a
compact RICH detector. Presently both options are being developed
for review by the DoE and the BNL-PAC. An additional forward
spectrometer is being considered for either option. A layout for the
forward components can be found in the Expression of Interest
\cite{r2d-eoi}.

In the present cost estimate about 40\% of the final cost is offset
by using existing high energy physics components. Negotiations with
the existing high energy physics collaborations have begun in order
to transition the necessary components after decommissioning.

\section{Summary}

The surprising discovery of a liquid-like phase of matter above the
critical parameters for a phase transition to a quark-gluon plasma
needs to be investigated in detail in order to understand
hadronization and chirality in QCD. This is a program which will
enable us to experimentally test the theory of asymptotic freedom
\cite{wilczek}. The question of how matter generates its mass is
fundamental to our understanding of the universe. The properties of
the created partonic phase have recently been compared to quantum
black holes \cite{kharzeev} and I think it is interesting to note,
that these properties are distinctly non-hadronic, which for the
first time enables us to probe the transition from partons to
hadrons from an initially thermalized partonic phase. Detailed
measurements of the degrees of freedom near T$_{c}$, the running of
the strong coupling constant above T$_{c}$, the quark separation of
fragmentation contributions to final hadrons and the determination
of chiral symmetry restoration make for an exciting new program at
RHIC-II.

\section*{Acknowledgements}

I thank John Harris, Andre Peshier, Rob Pisarski, Derek Teaney and
Simon Albino for useful discussions.


\vfill\eject
\end{document}